\documentclass{tlp}
\usepackage{aopmath}             %--- INSERIRE
\usepackage{graphics}
\usepackage{latexsym}

\newcommand{\nop}[1]{}
\newtheorem{definition}{Definition}
\newtheorem{example}{Example}

\newcommand{\vs}{\vspace*{-1mm}\noindent}

\newcommand{\PP}{\mbox{$P$}}
\newcommand{\tuple}[1]{\langle#1\rangle}

\newcommand{\Or}{\vee}

\newcommand{\Pol}{{\rm P}}
\newcommand{\NP}{{\rm NP}}

\newcommand{\CONP}{\mbox{\rm co-}\NP }

\newcommand{\SigmaP}[1]{{\Sigma}_{#1}^{P}}
\newcommand{\PiP}[1]{{\Pi}_{#1}^{P}}

\def\gp{{$LP$}}
\def\gp{{P}}
\def\F{\mbox{$\cal F$}}  % minimal model semantics
\def\M{\mbox{$\cal M$}}  % minimal model semantics
\def\MM{\mbox{$\cal MM$}}  % minimal model semantics
\def\SM{\mbox{$\cal SM$}}   % stable model semantics
\def\MF{\mbox{$\cal MF$}}  % minimal founded model semantics
\def\PM{\mbox{$\cal PM$}}  % perfect model semantics
\def\BP {B_{\gp}}

\def\SS{{\bf S}}
\def\WW{{\bf W}}
\def\XX{{\bf X}}
\def\YY{{\bf Y}}

\def\<{\langle}
\def\>{\rangle}

\def\DS{\mbox{$D\!S$}}
\def\PD{\PP\!_{D}}
\def\DB{D\!B}
\def\RR{{\bf R}}
\def\DD{{\bf D}}
\def\Q{\mbox{$Q$}}
\def\QQ{{\bf Q}}

\def\POSS{{\exists}}
\def\EPOSS{{E \! X \! P}^\POSS}
\def\CERT{{\forall}}
\def\ECERT{{E \! X \! P}^\CERT}

\begin{document}
\bibliographystyle{acmtrans}

\date{}

\title[Minimal Founded Semantics]
{Minimal Founded Semantics for Disjunctive \\  Logic Programs and
Deductive Databases\thanks{ A preliminary version of this paper
has  been presented at the LPNMR'99 conference (Greco, 1999). Work
partially supported by the Murst projects ``DataX" and ``D2I". The
third author is also supported by ISI-CNR.} }
\author[F. Furfaro, G. Greco and S. Greco]
{Filippo Furfaro, Gianluigi Greco and Sergio Greco   \\ \ \\
DEIS \\
Universit\`a della Calabria \\
87030 Rende, Italy \\
\{filippo.furfaro,gianluigi.greco,greco\}@deis.unical.it}

\pagerange{\pageref{firstpage}--\pageref{lastpage}}
\volume{\textbf{10} (3):}
\jdate{March 2002}
\setcounter{page}{1}
\pubyear{2002}

\maketitle
\label{firstpage}

\begin{abstract}
In this paper, we propose a variant of stable model semantics for
disjunctive logic programming and deductive databases. The
semantics, called \emph{minimal founded}, generalizes stable model
semantics for normal (i.e. non disjunctive) programs but differs
from disjunctive stable model semantics (the extension of stable
model semantics for disjunctive programs). Compared with
disjunctive stable model semantics, minimal founded semantics
seems to be more intuitive, it gives meaning to programs which are
meaningless under stable model semantics and is no harder to
compute. More specifically, minimal founded semantics differs from
stable model semantics only for disjunctive programs having
constraint rules or rules working as constraints. We study the
expressive power of the semantics and show that for general
disjunctive datalog programs it has the same power as disjunctive
stable model semantics.
\end{abstract}

\section{Introduction}

Several different semantics have been proposed for normal and
disjunctive logic programs. Stable model semantics, first proposed
for normal (i.e. disjunction free) programs, has been subsequently
extended to disjunctive programs. For normal programs, stable
model semantics has been widely accepted since it captures the
intuitive meaning of programs and, for stratified programs it
coincides with perfect model semantics which is the standard
semantics for this class of programs \cite{AptBla*88,Prz88,PrzPrz88,VanRos91}.
For positive programs,
stable model semantics coincides with minimal model semantics which
is the standard semantics for positive disjunctive programs.

For general disjunctive programs several semantics have been
proposed. We mention here the \emph{generalized closed world
assumption} (GCWA) \cite{mink-82}, the \emph{weak generalized
closed world assumption} (WGCWA) \cite{RajLob*89,LMR92}, the
\emph{possible model semantics} \cite{SakIno94}, the \emph{perfect
model semantics} \cite{Prz91}, particularly suited to stratified
programs, the \emph{disjunctive well-founded semantics}
\cite{Ros89}, the \emph{disjunctive stable model semantics}
\cite{GelLif91,Prz91} and the \emph{partial stable model
semantics} \cite{Prz91,EitLeoSac98}.

Disjunctive stable model semantics is widely accepted since i) it
gives a good intuition of the meaning of programs, ii) for normal
programs it coincides with stable model semantics and for
stratified (resp. positive) programs it coincides with the perfect
(resp. minimal) model semantics. However, disjunctive stable model
semantics has some drawbacks. It is defined for a restricted class
of programs and there are several reasonable programs which are
meaningless, i.e. they do not have stable models.

\subsection*{Motivating examples}

The following examples present some programs whose intuitive
meaning is not captured by disjunctive stable model semantics.

\begin{example}\label{party}
Consider the following simple disjunctive program $P_1$
\[
\begin{array}{l}
a \Or b \Or c  \leftarrow  \\
\leftarrow  \neg a \\
\leftarrow  \neg b
\end{array}
\]
where the second and third rules are constraints, i.e. rules which
are satisfied only if the body is false, which can be rewritten
into equivalent normal rules.\footnote{ A constraint rule of the
form $\leftarrow b_1,\ldots,b_k$ can be rewritten under total
semantics (i.e. a two value semantics where every atom is either
true or false) as $p(X) \leftarrow b_1,\ldots,b_k, \neg p(X)$ where
$p$ is a new predicate symbol and $X$ is the list of all distinct
variables appearing in the source rule.} $P_1$ has a unique
minimal model $M_1 = \{ a, b \}$ but $M_1$ is not
stable.~\hfill~$\Box$
\end{example}

Thus, under stable model semantics the above program is
meaningless. However, the intuitive meaning is captured by the
unique minimal model since the constraints force more than one
atom to be inferred from the disjunctive rule. The next example
presents a real life situation that can be easily modeled by means
of a disjunctive program.

\begin{example}\label{example2-introduction}
Consider the Internet structure where every computer in the
network makes use of a primary DNS (Domain Name Server) for
resolving names associated to IP addresses; moreover if the
primary server fails, a secondary (supplementary) DNS is searched.
So, an address cannot be resolved if both primary and secondary
DNSs are not reachable. An interesting task could be the
identification of a minimal set of servers that ensures the
connectivity of a set of computers. This task can be formalized by
the following disjunctive program:

\begin{eqnarray*}
active(D_1) \Or active(D_2) \leftarrow dns(C,D_1,D_2)
\end{eqnarray*}

\noindent where $active(D)$ means that $D$ is a working DNS,
$dns(C,D_1,D_2)$ means that $C$ is a computer with $D_1$ and $D_2$
as primary and secondary DNSs. Assuming that $dns$ is a relation
of our database, it is easy to see that this program has minimal
(stable) models (under the disjunctive stable model semantics) and
that each stable model corresponds to the set of working DNSs.

Now suppose that we are looking for a set of active DNSs
containing both $d_1$ and $d_2$; this situation can be modeled
by adding to the program the following constraint:

\begin{eqnarray*}
\leftarrow \neg active(d_1) \\
\leftarrow \neg active(d_2)
\end{eqnarray*}

\noindent Under this hypothesis, if there is a computer $c$ with
$d_1$ and $d_2$ as primary and secondary DNSs (i.e. there is a
fact $dns(c,d_1,d_2)$ in the database), the program has a minimal
model containing $active(d_1)$ and $active(d_2)$; but this model
is not stable. Thus, under stable model semantics this program is
meaningless, even though its intuitive meaning is captured by the
minimal model.~\hfill~$\Box$
\end{example}

For a better understanding of this problem, consider now the
formalization in terms of logic programming of the 3SAT problem.

\begin{example}\label{ex}
The 3SAT problem in which clauses consist of exactly 3 literals
can be expressed by the following three rules:

\[
\begin{array}{ll}
\hspace*{-3mm} val(X,true) \Or val(X,false) \leftarrow var(X)  \\
\hspace*{-3mm} \leftarrow val(X,true), val(X,false) \vspace*{2mm} \\
\hspace*{-3mm} val(X,Vx) \Or val(Y,Vy) \Or val(Z,Vz) \leftarrow &
               occur(C,X,Vx), occur(C,Y,Vy),  \\
               &  occur(C,Z,Vz)
\end{array}
\]

The first two rules state that the value of each literal must be
either \emph{true} or \emph{false}. In the third rule a predicate
$occur(C,X,Vx)$ checks if the literal $X$ occurs in the clause
$C$; the value of $Vx$ is \emph{true} (resp. \emph{false}) if $X$
occurs positively (resp. negatively) in $C$. The set of clauses is
described by means of the database predicate $occur$. For
instance, the clause $c_1 = x_1 \Or x_2 \Or \neg x_3$ is defined
by the three facts $occur(c_1,x_1,true)$, $occur(c_1,x_2,true)$
and $occur(c_1,x_3,false)$. For the sake of simplicity, we are
assuming that all clauses consist of exactly three literals. Thus,
the third rule above states that for each clause, at least one of
its literals must be satisfied.

The above program, for an assigned set of input clauses, has a
number of models corresponding to all the truth assignments that
satisfy all the clauses; so asking for one model is equivalent to
solving the 3SAT problem.

Now suppose that one wants to find a solution in which two
variables $x_1$ and $x_2$ are both true: this situation is modeled
as usual by  means of the following two constraints:

\begin{eqnarray*}
\leftarrow \neg val(x_1,true) \\
\leftarrow \neg val(x_2,true)
\end{eqnarray*}

If there is no clause in which both $x_1$ and $x_2$ appear
positively, the program still solves the 3SAT problem with
constraint; but if there is such a clause then the program has no
minimal stable model because the constraint forces more than one
atom to be inferred from a disjunctive rule, and the minimal model
becomes not stable.~\hfill~$\Box$
\end{example}

Observe that the first two clauses in the program of the above
example can be rewritten into the following normal rules

\begin{eqnarray*}
val(X,true) \leftarrow var(X),\ \neg val(X,false) \\
val(X,false) \leftarrow var(X),\ \neg val(X,true)
\end{eqnarray*}

\noindent
since they are used to define a partition of the
relation $var$ and the constraint defined by the second rule is
used to force exclusive disjunction. Observe also that the
constraints $\leftarrow \neg val(x_1,true)$ and $\leftarrow \neg
val(x_2,true)$ are used to infer, if possible, the atoms
$val(x_1,true)$ and $val(x_2,true)$. These constraints cannot be
replaced by the two facts $val(x_1,true) \leftarrow$ and
$val(x_2,true) \leftarrow$ since by doing so we assert that $x_1$
and $x_2$ are true whereas the constraints are used to force the
semantics to infer, if possible, that $x_1$ and $x_2$ are true.

Intuitively, the problem with stable model semantics is that in
some cases the inclusive disjunction is interpreted as exclusive
disjunction. This is an old problem first noticed in
\cite{RosTop88} who proposed an alternative rule, called
{disjunctive database rule} (DDR), to infer negative information.
DDR is equivalent to the weak generalized closed world assumption
\cite{RajLob*89}, an extension of the generalized closed world
assumption proposed in \cite{mink-82}.

In this paper we try to
conjugate minimality of models and inclusive disjunction by
presenting a new semantics, called {\em minimal founded}, which
overcomes some drawbacks of disjunctive stable model semantics and
gives meaning to a larger class of programs by interpreting disjunction
in a more liberal way.

\subsection*{Contributions}

The main contributions of the paper are the following:
\begin{itemize}
\item
\vs We introduce a  semantics for disjunctive programs. The
proposed semantics seems to be more intuitive than stable model
semantics and it gives meaning to programs which are meaningless
under disjunctive stable model semantics.
\item
We show that the new semantics coincides with disjunctive stable
model semantics for normal and positive programs.
\item
We formally define the expressive power and complexity of the
proposed semantics for datalog programs and we show that it has
the same expressive power and complexity of disjunctive stable
model semantics.
\end{itemize}

As a consequence, the proposed semantics differs from stable model
semantics only for programs containing both disjunctive rules and
negation.

Although the full expressive power of disjunctive datalog can be
reached by only considering stratified programs, the natural way
to express NP problems and problems in the second level of the
polynomial hierarchy ($\Sigma_p^2$ and $\Pi_p^2$ problems) is to
use the \emph{guess-and-check} technique, where the \emph{guess}
part is expressed by means of disjunctive rules and the
\emph{check} part is expressed by means of constraints (i.e.
unstratified rules) \cite{EitLeo*98}. However, as shown by the
previous examples, there are several interesting programs whose
intuitive semantics is not captured by stable models. Thus, the
problem of defining an intuitive semantics for disjunctive datalog
is still an interesting topic.

We point out that the aim of this paper is not the introduction of
a more powerful semantics but only the definition of a semantics
which gives an intuitive meaning to a larger class of programs. In
the same way, disjunctive stable models do not increase the
expressive power of stratified disjunctive datalog under the
perfect model semantics, but just give semantics to a larger class
of programs.

\subsection*{Organization of the paper}

The rest of the paper is organized as follows. Section
\ref{preliminaries} presents preliminaries on disjunctive datalog,
minimal and stable model semantics. Section 3 introduces the {\em
minimal founded} semantics. Its relation with minimal model
semantics and stable model semantics is investigated. Section 4
presents results on the expressive power and complexity of minimal
founded semantics. Finally, Section 5 presents our conclusions.

%==========================================================

\section{Preliminaries}\label{preliminaries}

% \subsection{Disjunctive Logic Programs}

A {\em (disjunctive datalog) rule} $r$ is a clause of the form
\[
A_1 \Or \cdots \Or A_k \leftarrow B_1,\ldots, B_m,
    \neg\, C_1,\ldots, \neg\, C_n, \ \ \ \ \ \ \ \ \ \ k+m+n>0.
\]
where $A_1,\ldots ,A_k,B_1,\ldots ,B_m,C_1,\ldots ,C_n$ are atoms
of the form $p(t_1,..., t_h)$, $p$ is a \emph{predicate} of arity
$h$ and the terms $t_1,...,t_h$ are either constants or
variables. The disjunction $A_1\Or\cdots\Or A_k$ is the {\em head}
of $r$, while the conjunction $B_1,\ldots,B_m,\neg C_1,\ldots,\neg
C_n$ is the {\em body} of $r$. Moreover, if $k = 1$ we say that
the rule is normal, i.e. not disjunctive.

We denote by $Head(r)$ the set $\{A_1 ,\ldots, A_k \}$ of the head
atoms, and by $Body(r)$ the set $\{ B_1,\ldots
,B_m,\neg C_1, \ldots,\neg C_n \}$ of the body literals. We
often use upper-case letters, for example $L$, to denote literals.
As usual, a literal is an atom $A$ or a negated atom $\neg A$; in
the former case, it is positive, and in the latter negative.
Two literals $L_1$ and $L_2$ are {\em complementary} if $L_1=A$ and
$L_2 = \neg A$, for some atom $A$. For a literal $L$, $\neg L$ denotes
its complementary literal, and for a set $S$ of literals, $\neg S
= \{\neg L~|~L \in S\}$.
Moreover, $Body^{+}(r)$ and $Body^{-}(r)$ denote the set of positive
and negative literals occurring in $Body(r)$, respectively.

A {\em (disjunctive) logic program}  is a finite set of rules. A
$\neg$-free (resp. $\Or$-free) program is called {\em positive}\/
(resp. {\em normal}).  A term, (resp.\ an atom, a literal, a rule
or a program) is {\em ground}\/ if no variables occur in it. In
the following we also assume the existence of rules with empty
head, called \emph{denials}, which define
constraints\footnote{Under total semantics}, i.e. rules which are
satisfied only if the body is false.

The {\em Herbrand Universe} $U_{P}$ of a program $\gp$ is the set
of all constants appearing in $\gp$, and its {\em Herbrand Base}
$B_{P}$ is the set of all ground atoms constructed from the
predicates appearing in $\gp$ and the constants from $U_{P}$.  A
rule $r'$ is a {\em ground instance}\/ of a rule $r$, if $r'$ is
obtained from $r$ by replacing every variable in $r$ with some
constant in $U_{P}$. We denote by $ground(\gp)$ the set of all
ground instances of the rules in $\gp$.

Given a program $\PP$ and two predicate symbols (resp. ground
atoms) $p$ and $q$, we write $p \rightarrow q$ if there exists a
rule where $q$ occurs in the head and $p$ in the body or there
exists a predicate (resp. ground atom) $s$ such that $p
\rightarrow s$ and $s \rightarrow q$. If $p \rightarrow q$ then we
say that $q$ {\em depends on} $p$; also we say that $q$ {\em
depends on} any rule where $p$ occurs in the head. A predicate
(resp. ground atom) $p$ is said to be recursive if $p \rightarrow
p$.

An interpretation of $\gp$ is any subset of $\BP$. The value of a
ground atom $L$ w.r.t. an interpretation $I$, $value_I(L)$, is
$true$ if  $L\in I$ and $false$ otherwise. The value of a ground
negated literal $\neg L$ is $\neg value_I(L)$. The truth value of
a conjunction of ground literals $C = L_1,\ldots,L_n$ is the
minimum over the values of the $L_i$, i.e., $value_I(C) =
min(\{value_I(L_i)\ |\ 1 \leq i \leq n \})$, while the value
$value_I(D)$ of a disjunction $D = L_1 \vee ... \vee L_n$ is their
maximum, i.e., $value_I(D) = max(\{value_I(L_i)\ |\ 1 \leq i \leq
n \})$; if $n=0$, then $value_I(C) = true$ and $value_I(D) =
false$.  Finally, a ground rule $r$ is {\em satisfied} by $I$ if
$value_I(Head(r)) \geq value_I(Body(r))$. Thus, a rule $r$ with
empty body is satisfied by $I$ if $value_I(Head(r)) = true$
whereas a rule $r'$ with empty head is satisfied by $I$ if
$value_I(Body(r')) = false$. An interpretation $M$ for $\gp$ is a
model of $\gp$ if  $M$ satisfies each rule in $ground(\gp )$. The
set of all models of $\PP$ will be denoted by $\M(P)$.

Minker proposed in \cite{mink-82} a model-theoretic semantics for
a positive program $\gp$, which assigns  to $\gp$ the set of its
{\em minimal models} $\MM(\gp)$, where a model $M$ for $\gp$ is
minimal, if no proper subset of $M$ is a model for $\gp$.
Accordingly, the program  \mbox{$\gp =\{ a\Or b \leftarrow\}$} has
the two minimal models $\{a\}$ and $\{b\}$, i.e. $\MM(\gp )= \{\
\{a\},\ \{b\}\ \}$. The more general {\em disjunctive stable model
semantics} also applies to programs with (unstratified) negation
\cite{GelLif91,Prz91}. Disjunctive stable model semantics
generalizes stable model semantics, previously defined for normal
programs \cite{GelLif88}.

\begin{definition}
Let $P$ be logic program $P$ and let $I$ be an interpretation for
$P$, $\frac{P}{I}$ denotes the ground positive program derived
from $ground(\PP)$
\begin{enumerate}
\item
by removing all rules that contain a negative literal $\neg a$ in
the body and $a \in I$, and
\item
by removing all negative literals from the remaining rules.
\end{enumerate}
An interpretation $M$ is a (disjunctive) stable model of $\PP$ if
and only if $M \in \MM(\frac{P}{M})$.~\hfill~$\Box$
\end{definition}

For general $\gp$, the stable model semantics assigns to $\gp$ the
set $\SM(\gp)$ of its {\em stable models}. It is well known that
stable models are minimal models (i.e. $\SM(P) \subseteq \MM(P)$)
and that for negation-free programs minimal and stable model
semantics coincide (i.e. $\SM(P) = \MM(P)$).

An extension of the perfect model semantics for stratified datalog
programs to disjunctive programs has been proposed in
\cite{Prz91}.

A disjunctive datalog program $P$ is said to be {\em locally
stratified} if there exists a decomposition $S_1,...,S_\omega$  of
the Herbrand base such that for every (ground instance of a)
clause
\[
A_1 \vee ... \vee A_k \leftarrow B_1,\ldots,B_m, \neg\,
C_1,\ldots,\neg\,C_n
\]
in $P$, there exists an $l$, called level of the clause, so that:
\begin{enumerate}
\item $\forall i \leq k$ $stratum(A_i) = l$,
\item $\forall i \leq m$ $stratum(B_i) \leq l$, and
\item $\forall i \leq n$ $stratum(C_i) < l$.
\end{enumerate}
where $stratum(A) = i$ iff $A \in S_i$.

The set of clauses in $ground(P)$ having level $i$ (resp. $ \leq
i$) is denoted by $P_i$ (resp. $P_i^*$). Any decomposition of the
ground instantiation of a program $P$ is called local
stratification of $P$.

The preference order on the models of $P$ is defined as follows:
$M \prec N$ iff $M \neq N$ and for each $a \in M - N$ there exists
a $b \in N-M$ such that $stratum(a) > stratum(b)$. Intuitively,
$stratum(a) > stratum(b)$ means that $a$ has higher priority than
$b$.

\begin{definition}
Let $P$ be a locally stratified disjunctive datalog program. A
model $M$ for $P$ is {\em perfect} if there is no model $N$ such
that $N \prec M$. The collection of all perfect models of $P$ is
denoted by $\PM(P)$. \hfill $\Box$
\end{definition}

Consider for instance the program consisting of the clause $a \vee
b \leftarrow \neg\, c$. The minimal models are $M_1 = \{ a \}$,
$M_2 = \{ b \}$ and $M_3 = \{ c \}$. Since $stratum(a) >
stratum(c)$ and $stratum(b) > stratum(c)$, we have that $M_1 \prec
M_3$ and $M_2 \prec M_3$. Therefore, only $M_1$ and $M_2$ are
perfect models.

Notice that $M \subset N$ implies $M \prec N$; thus, for locally
stratified $P$, $\PM(P) \subseteq \MM(P)$. For positive $P$,
$\MM(P) = \PM(P)$ and for stratified $P$, $\PM(P) = \SM(P)
\subseteq \MM(P)$. The computation of the perfect model semantics
of a program $P$ can be done by considering a decomposition
$(P_1,\ldots,P_\omega)$ of $ground(P)$ and computing the minimal
models of all subprograms, one at time, following the linear order
\cite{FerMin91,greco-98,greco-99a}. In the decomposition $(P_1,\ldots,P_\omega)$, for
each $P_i$ and for each rule $r$ of $P_i$, if $A \in Head(r)$ and
$B \in Body^{+}(r)$ (resp. $B \in Body^{-}(r)$) then $B$ does not
appear in the head of any rule of $P_j$ with $j > i$ (resp. $j
\geq i$).

%========================================================================

\section{Minimal Founded Semantics}

In this section we introduce a new semantics for disjunctive
programs.

\begin{definition}
Let $P$ be a positive disjunctive program and let $M$ be an
interpretation. Then,

\[ S_P(M) = \{ a \in B_P | \exists r \in ground(P)  \wedge a \in Head(r) \wedge Body(r) \subseteq M \} \]

\noindent $S_P^\omega(\emptyset)$ denotes the least fixpoint of
the operator $S_P$. \hfill $\Box$
\end{definition}

The operator $S_P$ extends the classical immediate consequence
operator $T_P$ to disjunctive programs by replacing head
disjunctions with conjunctions. It is obvious that the operator
$S_P$, for positive $P$, is monotonic and continuous and,
therefore, it admits a least fixpoint.

\begin{definition}[Minimal Founded Semantics]
Let $P$ be a disjunctive program and let $M$ be a model for $P$.
Then, $M$ is a {\em founded} model if it is contained in
$S_{\frac{P}{M}}^\omega(\emptyset)$. $M$ is said to be {\em
minimal founded} if it is a minimal model of $P$ and it is also
founded. The collection of all minimal founded models of $P$ is
denoted by $\MF(P)$. ~\hfill~$\Box$
\end{definition}

For any program $P$, the set of founded models of $P$ will be
denoted by $\F(P)$.

\begin{example}
The program $P_1$ of Example 1 has a unique minimal model $M_1 =\{
a, b \}$ which is also founded since it is the fixpoint of
$S_{\frac{P_1}{M_1}}$. Observe that the interpretation $N_1 = \{
a, b, c \}$ is a founded model for $P_1$ but it is not minimal
since $M_1 \subset N_1$. \hfill $\Box$
\end{example}

\begin{fact}\label{fact-minimal-founded}
Let $P$ be a disjunctive datalog program. Then, $\MF(P) \subseteq
\MM(P)$.

\noindent {\bf Proof.} By definition of minimal founded model.
\hfill $\Box$
\end{fact}

The following example presents a disjunctive program where stable
and minimal founded semantics coincide.

\begin{example}\label{MF-example}
Consider the following simple disjunctive program
$P_{\ref{MF-example}}$
\[
\begin{array}{l}
a \Or b \Or c  \leftarrow  \\
a \leftarrow  \neg b, \neg c \\
b \leftarrow  \neg a \\
c \leftarrow  \neg a
\end{array}
\]
This program has two stable models $M_{\ref{MF-example}} = \{ a
\}$ and $N_{\ref{MF-example}} = \{ b, c \}$ which are also minimal
founded. \hfill $\Box$
\end{example}

Moreover, for general programs containing both disjunction and
negation, stable and minimal founded semantics do not coincide.
The relation between the two semantics is given by the following
result.

\begin{theorem}\label{theorem-stable-mfounded}
Let $P$ be a disjunctive program. Then, $\SM(P) \subseteq \MF(P)$.
\end{theorem}
\vspace{-2mm} \noindent {\bf Proof.} Since stable models are
minimal models, we have to show that any stable model $M$ of $P$
is founded, i.e. $M \subseteq S_{\frac{P}{M}}^\omega(\emptyset)$.
Since $\frac{P}{M}$ is negation-free, every minimal model of
$\frac{P}{M}$ is contained in $S_{\frac{P}{M}}^\omega(\emptyset)$.
Thus, $M$ is founded and, consequently, $\SM(P) \subseteq \MF(P)$.
\hfill $\Box$

\vspace*{3mm} Therefore, for every disjunctive program $P$,
$\SM(P) \subseteq \MF(P) \subseteq \MM(P)$. Moreover, as shown by
the previous examples, there are programs where the containment is
strict, i.e. there are programs, such as the ones presented in the
Introduction, having minimal founded models which are not stable.

\begin{corollary}
Let $P$ be a positive disjunctive datalog program. Then, $\MM(P) =
\MF(P)$.

\noindent {\bf Proof.} From Theorem \ref{theorem-stable-mfounded}
$\SM(P) \subseteq \MF(P)$. Moreover, by definition $\MF(P)
\subseteq \MM(P)$. \ Since for positive programs \ $\SM(P) =
\MM(P)$, we conclude that \ $\MF(P) = \MM(P)$. \hfill $\Box$
\end{corollary}

The following result states that for disjunction-free programs,
stable model semantics and minimal founded semantics coincide.

\begin{proposition}\label{proposition-minimal-founded}
Let $P$ be a normal datalog program. Then, $\SM(P) = \MF(P)$.
\end{proposition}
{\bf Proof}. Generally, $\SM(P) \subseteq \MM(P)$. Thus we have to
show that every minimal founded model is also stable. Since for
every normal program $P$ and any interpretation $M$ of $P$, the
operators $T_{\frac{P}{M}}$ and $S_{\frac{P}{M}}$ coincide, we
have that every minimal  founded model $M$ of $P$ is equal to
$T^\omega_{\frac{P}{M}}(\emptyset)$. ~\hfill~$\Box$

\vspace{3mm} The following example presents another case of a
program which is meaningless under stable model semantics but has
minimal founded models.

\begin{example}\label{Unstratified-Example}
Consider the program $P_{\ref{Unstratified-Example}}$
\[
\begin{array}{l}
a \vee b \vee c \leftarrow \\
a \leftarrow \neg b \\
b \leftarrow \neg c \\
c \leftarrow \neg a
\end{array}
\]
From the first rule we have that a subset of $\{ a, b, c \}$ must
be selected whereas the last three rules state that at least two
atoms among $a$, $b$ and $c$ must be true. The program has three
minimal founded models, $M_{\ref{Unstratified-Example}} = \{ a, b
\}$, $N_{\ref{Unstratified-Example}} = \{ b, c \}$ and
$H_{\ref{Unstratified-Example}} = \{ a, c \}$, but none of them is
stable. \hfill $\Box$
\end{example}

It is worth noting that a disjunctive program $P$ may have no, one
or several minimal founded models. In the previous example we have
presented programs which are meaningless under the stable model
semantics which have minimal founded models (those presented in
the Introduction) and a program where stable and minimal founded
semantics coincide. The following example presents a program which
has stable models but the stable and minimal founded semantics do
not coincide.

\begin{example}\label{Stratified-Example}
Consider the program $P_{\ref{Stratified-Example}}$
\[
\begin{array}{cll}
eat \Or drink & \leftarrow & \\
eat & \leftarrow & \\
thirsty & \leftarrow & \neg drink
\end{array}
\]
This program has two minimal founded models
$M_{\ref{Stratified-Example}} = \{ eat, thirsty \}$ and
$N_{\ref{Stratified-Example}} = \{ eat, drink \}$, but only
$M_{\ref{Stratified-Example}}$ is stable. \hfill $\Box$
\end{example}

We now introduce a different characterization of the minimal
founded semantics which permits us to better understand the
relationship between stable and minimal founded semantics.

\begin{definition}\label{transformation-definition}
Let $P$ be a disjunctive program and let $M$ be an interpretation.
Then, $P^M$ denotes the program derived from $ground(P)$ by
deleting for each rule
\[
r: A_1 \vee \cdots \vee A_k \leftarrow B_1,\ldots,B_m,\neg C_1,\ldots,
\neg C_n
\]
every $A_i \not\in M$. \hfill $\Box$
\end{definition}

\begin{proposition}
Let $P$ be a disjunctive program and let $M$ be an interpretation.
Then $M \in \MF(P)$ if and only if $M \in \MF(P^M)$.
\end{proposition}

\vspace{-2mm} \noindent {\bf Proof.} $M$ is a minimal founded
model of $P$ iff it is a minimal founded model of $P'=ground(P)$.
$M$ is a minimal model for $P'$ if and only if it is a minimal
model of $P^M$ since we delete from $P'$ head atoms which are
false in $M$. Moreover, if an atom can be inferred in
$\frac{P'}{M}$ it can also be  inferred in $\frac{P^M}{M}$ and
vice-versa, i.e. $\F(\frac{P'}{M}) = \F(\frac{P^M}{M})$.
% $S^\omega_{\frac{P'}{M}}(\emptyset) = S^\omega_{\frac{P^M}{M}}(\emptyset)$.
Therefore, $M$ is a minimal founded model for $P'$ iff it is a
minimal founded model for $P^M$. \hfill $\Box$

\vspace{3mm} Observe that the program $P^M$ consists of standard
rules whose head is not empty and denials (rules with empty head).
Thus, in the following we shall denote with $P_S^M$ the set of
standard rules of $P^M$ whose head is not empty and with $P_D^M$
the set of denial rules of $P^M$.

\begin{theorem}\label{Theorem2}
Let $P$ be a disjunctive datalog program and $M$ a minimal model
for $P$. Then, $M \in \MF(P)$ if and only $M \in \F(P_S^M)$ and $M
\models P_D^M$.
\end{theorem}

\vspace{-2mm} \noindent {\bf Proof.} Clearly $M$ is a minimal
founded model for $P$ iff it is a minimal founded model for
$P'=ground(P)$.

We first prove that $M \in \MF(P')$ implies that $M \in \F(P_S^M)$
and $M \models P_D^M$. Let $P''$ be the subset of rules in $P'$
from which the rules in $P_D^M$ are derived. Every denial $r: \
\leftarrow B_1,...,B_m, \neg C_1,...,\neg C_n$, derived from a
rule $r'': A_1 \vee \cdots \vee A_k \leftarrow B_1,...,B_m, \neg
C_1,...,\neg C_n$, is satisfied in $M$ if and only if $r''$ is
also satisfied in $M$ because all atoms $A_1,...,A_k$ are false in
$M$. As $P_S^M = P' - P''$, if $M$ is a (minimal) founded model
for $P'$ it is also a founded model for $P_S^M$ since from the
rules in $P''$ it is not possible to infer any atom.

We now prove that if $M$ is a minimal model of $P$ such that $M
\in \F(P_S^M)$, then $M \in \MF(P')$. As $P_S^M \subseteq P'$, if
$M$ is a founded model for $P_S^M$ and is a minimal model for $P'$
it will be a minimal founded model for $P'$. It is obvious that if
$M$ is a minimal model of $P'$ every rule of $P'$ is satisfied.
\hfill $\Box$

\vspace*{3mm} It is important to note that in the ground program
there are rules which with respect to a given model act as
constraints forcing atoms to be true or false. In the following
example we reconsider the program $P_{\ref{Unstratified-Example}}$
of Example \ref{Unstratified-Example} containing rules which force
the selection of two atoms from the disjunctive rule.

\begin{example}\label{ABC-Example}
The program $P_{\ref{Unstratified-Example}}$ of Example
\ref{Unstratified-Example} \nop{
\[
\begin{array}{l}
a \vee b \vee c \leftarrow \\
a \leftarrow \neg b \\
b \leftarrow \neg c \\
c \leftarrow \neg a
\end{array}
\]
The program } admits three minimal founded models:
$M_{\ref{Unstratified-Example}} = \{ a, b \}$,
$H_{\ref{Unstratified-Example}} = \{ a, c \}$ and
$N_{\ref{Unstratified-Example}} = \{ b, c \}$. The program
$P^{M_{\ref{Unstratified-Example}}} =
(P_S^{M_{\ref{Unstratified-Example}}},P_D^{M_{\ref{Unstratified-Example}}})$
is
\[
\begin{array}{l}
a \vee b  \leftarrow \\
b \leftarrow  \\
\leftarrow \neg a
\end{array}
\]
where $P_S^{M_{\ref{Unstratified-Example}}}$ consists of the first
two rules and $P_D^{M_{\ref{Unstratified-Example}}}$ contains the
last rule. The only minimal model for
$P^{M_{\ref{Unstratified-Example}}}$ is
$M_{\ref{Unstratified-Example}}$; this model satisfies
$P_D^{M_{\ref{Unstratified-Example}}}$
 and is a founded model of $P_S^{M_{\ref{Unstratified-Example}}}$.

As the program $P_{\ref{Unstratified-Example}}$ is symmetric, we
have that also $H_{\ref{Unstratified-Example}}$ and
$N_{\ref{Unstratified-Example}}$ are minimal founded model of
$P_{\ref{Unstratified-Example}}$. \nop{---------------- The
program $P^{H_{\ref{Unstratified-Example}}}$ is
\[
\begin{array}{l}
a \vee c  \leftarrow \\
a \leftarrow  \\
\leftarrow \neg c
\end{array}
\]
The only minimal model for $P^{H_{\ref{Unstratified-Example}}}$ is
$H_{\ref{Unstratified-Example}}$; this model satisfies
$P_D^{H_{\ref{Unstratified-Example}}}$ and is a founded model of
$P_S^{H_{\ref{Unstratified-Example}}}$. The program
$P^{N_{\ref{Unstratified-Example}}}$ is
\[
\begin{array}{l}
b \vee c  \leftarrow \\
\leftarrow \neg b \\
c \leftarrow  \\
\end{array}
\]
The only minimal model for $P^{N_{\ref{Unstratified-Example}}}$ is
$N_{\ref{Unstratified-Example}}$. } \hfill $\Box$
\end{example}

Theorem \ref{Theorem2} shows the difference between minimal
founded and stable model semantics. In particular, given program
$P$ and a minimal model $M$ for $P$, $M$ is stable if $M$ is a
minimal model of $\frac{P_S^M}{M}$ and $M$ satisfies $P_D^M$
whereas $M$ is a minimal founded model if $M$ is a model of
$\frac{P_S^M}{M}$  and $M$ satisfies $P_D^M$. Thus, the main
difference between the two semantics is that the stable model
semantics asks for minimal models of $ground(P)$ which satisfy the
constraints $P_D^M$ and are also minimal for the subset of
standard rules $P_S^M$, whereas the minimal founded model asks for
minimal models of $ground(P)$ which satisfy the constraints
$P_D^M$ and are founded, i.e. their atoms are derivable from the
rules in $P_S^M$.

It is worth noting that the above result can be very useful in the
computation of the semantics of programs. Indeed, during the
computation of a model, from the assumption of the falsity of
atoms we derive constraints which further restrict the search
strategy \cite{LeoRul*97,EitLeo*98}.

%=========================================================================
%=========================================================================
%=========================================================================

\section{Expressive Power and Complexity}\label{results}

In this section we present some results about the expressive
power and the data complexity of minimal founded semantics for
disjunctive datalog programs  \cite{EitGot*97,EitLeoSac98,Sac97}.
We first introduce some preliminary definitions and notation,
and then present our results.

Predicate symbols are partitioned into the two sets of {\em base }
({\em EDB}) and {\em derived} ({\em IDB}) predicates. Base
predicate symbols correspond to database relations on a countable
domain $U$ and do not occur in the rule heads. Derived predicate
symbols appear in the head of rules. Possible constants in a
program are taken from the domain $U$.

A program $\PP$ has associated a relational database scheme $\DS_P
= \{r|$ $r$ is an EDB predicate symbol of $\PP\}$, thus EDB
predicate symbols are seen as relation symbols. A database $D$ on
$\DS_P$ is a set of finite relations, one for each $r$ in $\DS_P$,
denoted by $D(r)$. The set of all databases on $\DS_P$ is denoted
by $\DD_P$.

Given a database $D \in \DD_P$, $\PD$ denotes the following logic
program:
\[ \PD = \PP \cup \{r(t) \leftarrow \ |\ r \in \DS_P \wedge t \in D(r)\}. \]
\noindent The Herbrand universe $U_{P_D}$ is a finite subset of
$U$ and consists of all constants occurring in $\PP$ or in $D$
({\em active domain}). If $D$ is empty and no constant occurs in
$\PP$, then $U_{P_D}$ is assumed to be equal to $\{a\}$, where $a$
is an arbitrary constant in $U$.

\begin{definition}\label{def-query}
A {\em bound} {\em query} $\Q$ is a pair $\<\PP,g\>$, where $\PP$
is a disjunctive program and $g$ is a ground literal (the {\em
query goal}). \hfill $\Box$
\end{definition}

We use $XM$ as generic notation for a generic semantics. The
result of a query $\Q = \<\PP,g\>$ on an input database $D$ is
defined in terms of the $XF$ models of $\PP_D$, by taking either
the union of all models ({\em brave or possible inference},
$\POSS_{XF}$) or the intersection ({\em cautious or certain
inference}, $\CERT_{XF}$).

\begin{definition}\label{def-query1}
Given a program $\PP$ and a database $D$, a ground atom $g$ is
true, under the brave version of the $XF$ semantics, if there
exists an $XF$ model $M$ for $\PP_D$ such that $g \in M$.
Analogously, $g$ is true, under the cautious version of the $XF$
semantics, if $g$ is true in every $XF$ model. The set of all
queries is denoted by $\QQ$.~\hfill~$\Box$
\end{definition}

\begin{definition}\label{def-power}
Let $\Q =\<\PP,g\>$ be a bound query. Then the {\em database
collection} of $\Q$ w.r.t. the set of $XF$ models is:
\begin{enumerate}
\item[$(a)$]
{\em under the brave version of semantics}, the set of all
databases $D$ in $\DD_P$ such that $g$ is $true$ in $\PD$ under
the brave version of the $XF$ semantics; this set is denoted  by
$\EPOSS_{XF}(\Q)$;
\item[$(b)$]
{\em under the cautious version of semantics}, the set of all
databases $D$ in $\DD_P$ such that $g$ is $true$ in $\PD$ under
the cautious version of the $XF$ semantics; this set is denoted
by $\ECERT_{XF}(\Q)$.
\end{enumerate}
The {\em expressive power} of a given version (either brave or
cautious) of the $XF$ semantics is given by the family of the
database collections of all possible queries, i.e.,
$\EPOSS_{XF}[\QQ] = \{\EPOSS_{XF}(\Q)|\Q \in \QQ\}$ and
$\ECERT_{XF}[\QQ] = \{\ECERT_{XF}(\Q)|\Q \in \QQ\}$.~\hfill$\Box$
\end{definition}

The database collection of every query is indeed a generic set of
databases. A set $\DD$ of databases on a database  scheme $\DS$
with domain $U$ is ($W$-){\em generic} if there exists a finite
subset $W$ of $U$ such that for any $D$ in $\DD$ and for any
isomorphism $\theta$ on relations extending a permutation on
$U-W$, $\theta(D)$ is in  $\DD$ as well  \cite{ChaHar82,AHV94} ---
informally, all constants not in $W$ are not interpreted, and
relationships among them are only those explicitly provided by the
databases. Note that for a  query $Q = \<\PP,g\>$, $W$ consists of
all constants occurring in $\PP$ and in $g$. From now on, any
generic set of databases will be called a {\em database
collection}.

Following the {\em data complexity} approach of
\cite{ChaHar82,Var82} for which the query is assumed to be a
constant while the database is the input variable, the expressive
power coincides with the complexity class of the problem of
recognizing  the database collection of each query. The expressive power
of each semantics will be compared with database complexity
classes, defined as follows. Given a Turing machine complexity
class $C$ (for instance $\PP$ or $\NP$), a relational database
scheme $\DS$, and a database collection $\DD$ on $\DS$, $\DD$ is
{\em $C$-recognizable} if the problem of deciding whether $D$ is
in $\DD$ is in $C$. The {\em database complexity class} $\DB$-$C$
is the family of all $C$-recognizable database collections (for
instance, $\DB$-$\PP$ is the family of all database collections
that are recognizable in polynomial time). If the expressive power
of a given semantics coincides with a complexity class $\DB$-$C$,
we say that the given semantics captures (or expresses all queries
in) $\DB$-$C$.

Recall that the classes $\SigmaP{k}$, $\PiP{k}$ of the polynomial
hierarchy \cite{stoc-77} are defined by $\SigmaP{0} = \Pol$,
$\SigmaP{i+1} = \NP^{\SigmaP{i}}$, and $\PiP{i} =$
co-$\SigmaP{i}$, for all $i \geq 0$. In particular, $\PiP{0} =
\Pol$, $\SigmaP{1} = \NP$, and $\PiP{1} = \CONP$. Using Fagin's
Theorem \cite{fagi-74} and its generalization in \cite{stoc-77},
complexity and second-order definability are linked as follows.

\begin{fact}\label{prop:so=ph}  \cite{fagi-74,stoc-77}
A \ database \ collection \ $\DD$ \ over \ a \ scheme \ $DS$ \ is
\ in \ $DB$-$\SigmaP{k}$ (resp. $DB$-$\PiP{k}$), $k \geq 1$, iff
it is definable by a second-order formula $(\exists A_1)(\forall
A_2)\cdots(Q_k A_k)\Phi$ (resp. $(\forall A_1)(\exists
A_2)\cdots(Q_k A_k)\Phi$) on $DS$, where the $A_i$ are lists of
predicate variables preceded by alternating quantifiers and $\Phi$
is first-order.~\hfill~$\Box$
\end{fact}

The following example shows how a NP problem can be expressed by
means of a second order formula and how the formula can be
translated into a disjunctive datalog program under minimal
founded or stable model semantics.

\begin{example}
Consider the {\em graph kernel} problem defined as: {\em given a
directed graph $G=\<V,E\>$, does there exist a kernel for $G$},
i.e. is there a set $S \subseteq V$ of vertices such that both (i)
for each $i$ in $V-S$, there exists $j$ in $S$ for which the edge
$(j,i)$ is in $E$, and (ii) for each $i,j$ in $S$,  $(i,j)$ is not
in $E$?

\nop{
\begin{figure}
\centering \  \figK \caption{\em Directed Graph}\label{fig-kernel}
\end{figure}

\begin{example}
\em
The directed graph in Figure \ref{fig-kernel} has two
kernels: $\{1,2\}$ and $\{3,4\}$. \hfill $\Box$
\end{example}
}

We denote the set of all (finite) directed graphs with $\DD_{G}$,
the set of all  graphs in $\DD_{G}$ for which a kernel exists with
$\DD_G^K$, and  $\overline{\DD}_G^K=$ $\DD_{G}-\DD_G^K$. Any graph
is represented by a database on the database scheme $BD = \{ V, E
\}$, where $V$ and $E$ store its vertices and edges, respectively.

Consider the following second-order formula over $BD$:
\[
\exists S\, \forall x\, \{\, [\neg S(x) \wedge \exists y ( S(y)
\wedge E(y,x))] \vee [S(x) \wedge \forall y(S(y) \Rightarrow \neg
E(y,x)) ]\, \}
\]
Note that $V$ supplies the interpretation domain of the formula.
It is easy to see that a graph $G$ is in $\DD_G^K$ iff the formula
is satisfied by $G$. The above formula can be rewritten in the
following equivalent {\em Skolem normal format} for existential
second order formulas:

\[
\begin{array}{ll}
\exists S\,  \forall x_1,x_2\, \exists y \{ & [\neg S(x_1) \wedge
S(y) \wedge E(y,x_1)] \vee [S(x_1) \wedge \neg S(x_2)] \vee \\
&[S(x_1) \wedge S(x_2) \wedge \neg E(x_2,x_1)]\ \}
\end{array}
\]
This formula is then used to construct the following datalog
program:

\[
\begin{array}{l}
\tt r_1:\ s(W) \vee \hat{s}(W) \leftarrow  \\
\tt r_2:\ \leftarrow  s(W), \ \hat{s}(W). \\
\tt r_3:\ q(X_1,X_2) \leftarrow  \hat{s}(X_1), \ s(Y), \ e(Y,X_1). \\
\tt r_4:\ q(X_1,X_2) \leftarrow s(X_1), \ \hat{s}(X_2). \\
\tt r_5:\ q(X_1,X_2) \leftarrow s(X_1),  \ s(X_2), \ \neg e(X_2,X_1). \\
\tt r_6:\ g          \leftarrow \neg q(X_1,X_2).
\end{array}
\]

\noindent where $v$ and $e$ are EDB predicate symbols and $s$ and
$\hat s$ are used to define a partition of the database domain
(the Herbrand universe). Note that the rules ($r_3$)-($r_5$)
implement the three conjunctions in the above Skolem normal form
formula.

Let $G=\<V,E\>$ be a directed graph.  A minimal founded (or stable
model) is constructed as follows. \ The first two rules
non-deterministically select two disjoint subsets of $V$, say $S$
and $\hat{S}$ respectively. For each $x_1$ in $\hat{S}$, if there
exists a vertex $y$ in $S$ for which $(y,x_1)$ is in $G$ (i.e.
$x_1$ is connected to some vertex in $S$) then the third rule
makes $q(x_1,x_2)$ true for every $x_2$ in $V$. The fourth rule
makes $q(x_1,x_2)$ true for each $x_1$ in $S$ and for each $x_2$
in $\hat{S}$, and the fifth rule makes $q(x_1,x_2)$ true if both
$x_1$ and $x_2$ are in $S$ and the edge from $x_2$ to $x_1$ is not
in $G$. Note that $q(x_1,x_2)$ is derived to be true for every
$x_1,x_2$ in $V$ iff $S$ and $\hat{S}$ cover $V$ and $S$ is a
kernel. But $g$ is false iff for every $x_1,x_2$ in $V$,
$q(x_1,x_2)$ is true; so $g$ is false iff $S$ and $\hat{S}$ cover
$V$ and $S$ is a kernel.

For a graph for which a kernel exists, $g$ may be either true or
false. Moreover there exists at least one stable model which
selects a kernel and, therefore, makes $g$ false. For a graph
without kernels, $g$ is always true in every stable model. \hfill
$\Box$
\end{example}

%------------------------------------------------------------------------
%------------------------------------------------------------------------
%------------------------------------------------------------------------

It is well known that, under total stable model semantics,
disjunctive datalog captures the complexity classes $\Sigma_2^P$
and $\Pi_2^P$, respectively, under brave and cautious semantics
\cite{EitGot*97}, whereas plain datalog (i.e. datalog with
negation and without disjunction) captures the complexity classes
$\NP$ and $co\NP$, respectively, under brave and cautious
semantics \cite{MarTru91,Sch95}.

% \subsection{Minimal Founded Semantics}

We now present some results on the expressive power and data
complexity of the minimal founded semantics.

\begin{theorem}\label{theorem-complexity1}
Given a disjunctive program $\PP$, a database $D$ on $\DS_P$,  and
an interpretation $M$ for $\PD$, deciding whether $M$ is a minimal
founded model for $\PD$ is $co\NP$-complete.

\noindent {\bf Proof.} Let $M$ be an interpretation and consider
the complementary problem $\overline{\Pi}$: is it true that $M$ is
not a strongly founded model? $\overline{\Pi}$ is in $\NP$ since
we can guess an interpretation $N$ and verify in polynomial time
that either (i) $M$ is not a founded model for $\PD$ or (ii) $N$
is a model for $\PD$ and $N \subset M$. Hence the problem $\Pi$ is
in $co\NP$.

Moreover, deciding whether an interpretation $M$ for a positive
disjunctive program $\PD$ is a minimal model is $co\NP$-complete.
Since for positive programs minimal models are also founded, then,
deciding whether $M$ is minimal founded is $co\NP$-hard.
Therefore, deciding whether $M$ is a minimal founded model for
$\PD$ is $co\NP$-complete. \hfill $\Box$
\end{theorem}

Observe that, deciding whether an interpretation $M$ is a stable
model for $\PD$ is also $co\NP$-complete.

\begin{theorem}\label{theorem-certainty-power}
$\ECERT_{\cal MF}[\QQ]=\DB$-$\PiP{2}$.

\noindent {\bf Proof} We first prove that for any query
$\Q=\tuple{\PP,g}$ in $\QQ$, recognizing whether a database $D$ is
in $\ECERT_{\cal MF}(\Q)$ is in $\PiP{2}$. To this end, we
consider the complementary problem: is it true that $D$ {\em is
not} in $\ECERT_{\cal MF}(\Q)$? Now, $D$ is not in $\ECERT_{\cal
MF}(\Q)$ iff there exists a minimal founded model $M$ of $\PD$
such that $g \not\in M$. Following the line of the proof of
Theorem \ref{theorem-possibility-power}, we can easily see that
the latter problem is in $\SigmaP{2}$. Hence, recognizing whether
a database $D$ is in $\ECERT_{\cal MF}(\Q)$ is in $\PiP{2}$.

Let us now prove that every $\Pi_2^p$ recognizable database
collection $\DD$ on a database scheme $\DS$ is in $\ECERT_{\cal
MF}[\QQ]$. By Fact \ref{prop:so=ph}, $\DD$ is defined by a second
order formula of the form $\forall \RR^1 \exists \RR^2
\Phi(\RR^1,\RR^2)$. Using the usual transformation technique, the
above formula is equivalent to a second order Skolem form formula
$(\forall \SS^1)(\exists \SS^2)\Gamma(\SS^1,\SS^2)$, where
\[
\Gamma(\SS^1,\SS^2)=(\forall \XX)(\exists
\YY)(\Theta_1(\SS^1,\SS^2,\XX,\YY) \vee \dots \vee
\Theta_k(\SS^1,\SS^2,\XX,\YY)),
\]
$\SS^1$ and $\SS^2$ are two lists of, respectively, $m_1$ and
$m_2$ predicate symbols, containing all symbols in $\RR^1$ and
$\RR^2$, respectively. Consider the following program $\PP$:
\[\begin{array}{llll}
r_1:\ s_j^1(\WW_j^1) \Or \hat{s}_j^1(\WW_j^1) & \leftarrow &  & (1 \leq j \leq m_1) \\
r_2:\ s_j^2(\WW_j^2) \Or \hat{s}_j^2(\WW_j^2) & \leftarrow &  & (1 \leq j \leq m_2) \\
r_3:\ q(\XX) & \leftarrow & \Theta_i(\SS^1,\SS^2,\XX,\YY) \ \ \ \ \ \ & (1 \leq i \leq k)\\
r_4:\ g & \leftarrow & \neg q(\XX). & \\
r_5:\ g & \leftarrow & s_j^2(\WW_j^2), \hat{s}_j^2(\WW_j^2)  & (1 \leq j \leq m_2)\\
r_6:\ \hat{s}_j^2(\WW_j^2) & \leftarrow & g. & (1 \leq j \leq m_2) \\
r_7:\ s_j^2(\WW_j^2) & \leftarrow & g & (1 \leq j \leq m_2)
\end{array}
\]
where, \ intuitively, \ $\hat{s}_j^1(\WW_j^1)$ corresponds to \
$\neg s_j^1(\WW_j^1)$, \ $\hat{s}_j^2(\WW_j^2)$ \ corresponds to
$\neg s_j^2(\WW_j^2)$ and the rules of group $r_3$ defining $q$
are used to implement the disjunction of the above second order
formula. Observe that the guesses defined by the rules in the
groups $r_1$ and $r_2$ are used in the rules in the group $r_3$
defining the predicate $q$ and that the rules in the groups $r_5$,
$r_6$ and $r_6$ force $g$ to be false. Now it is easy to show that
the formula $(\forall \SS^1)(\exists \SS^2)\Gamma(\SS^1,\SS^2)$ is
valid if $g$ is false in all minimal founded models of $\PP$ (if
$g$ is true the last two sets of rules make the second group of
rules false). \hfill $\Box$
\end{theorem}

\begin{theorem}\label{theorem-possibility-power}
$\EPOSS_{\cal MF}[\QQ]=\DB$-$\SigmaP{2}$.

\noindent {\bf Proof.} We first prove that for any query
$\Q=\tuple{\PP,g}$ in $\QQ$, recognizing whether a database $D$ is
in $\EPOSS_{\cal MF}(\Q)$ is in $\SigmaP{2}$.  $D$ is in
$\EPOSS_{\cal MF}(\Q)$ iff there exists a minimal founded model
$M$ of $\PD$ such that $g \in M$. To check this, we may guess an
interpretation $M$ of $\PD$ and verify that $M$ is a minimal
founded model of $\PD$. The guess of the interpretation $M$ is
polynomial time. To check that $M$ is minimal founded we can ask
an $NP$ oracle. Therefore, recognizing whether a database $D$ is
in $\EPOSS_{\cal MF}(\Q)$ is in $\SigmaP{2}$.

Let us now prove that every $\Sigma_2^p$ recognizable database
collection $\DD'$ on a database scheme $BD$ is in $\EPOSS_{\cal
MF}[\QQ]$. We have that $\DD'$ is defined by a second order
formula of the form $\exists \RR^1 \forall \RR^2
\Phi'(\RR^1,\RR^2)$. By setting $\Phi(\RR^1,\RR^2) = \neg
\Phi'(\RR^1,\RR^2)$, we have that the formula  $\forall \RR^1
\exists \RR^2 \Phi(\RR^1,\RR^2)$ defines the database collection
$\DD$, where $\DD = \DD_{BD} - \DD'$ and $\DD_{BD}$ is the set of
all databases on $BD$. Consider the program $\PP$ and the query
$\Q=\<\PP,\neg g\>$ in the proof of Theorem
\ref{theorem-certainty-power}. In it we have  shown that a
database $D$ in $\DD_{BD}$ is in $\DD$ iff $D$ is in $\ECERT_{\cal
MF}(\Q)$; hence a database $D$ in $\DD_{BD}$ is in $\DD'$ iff $D$
is not in $\ECERT_{\cal MF}(\Q)$. But $D$ is not in $\ECERT_{\cal
MF}(\Q)$ iff there exists some stable model $M$ for which $g$ is
in $M$. It follows that $\DD' = \EPOSS_{\cal MF}(\Q')$ where
$\Q'=\<\PP,g\>$. \hfill $\Box$
\end{theorem}

Therefore, the expressive power of disjunctive datalog under
minimal founded and stable model semantics is the same.

\vspace{3mm} Data complexity is usually closely tied to expressive
power and, in particular, it provides an upper bound for the
expressive power \cite{EitGot93}.

\vspace{3mm}
In this section we have shown that minimal founded
semantics is complete for the second level of the polynomial
hierarchy. For the stable model semantics it has been shown that
for the class of head-cycle-free (hcf) the computation of a model
selected nondeterministically can be done in polynomial time and
checking if a ground atom belongs to a minimal model (resp. all
minimal models) is complete for the first level of the polynomial
hierarchy, i.e. NP-complete (resp. coNP-complete) \cite{Deckter}.
This result does not immediately apply  to the minimal founded
semantics since there could be rules which could force the
selection of more than one atom appearing in the head of a rule.
We conjecture that we have the same results for the class of
head-cycle-free programs where constraints do not force the
selection of more than one atom from the head of disjunctive
rules. It is possible to identify a syntactic class consisting of
hcf programs where after the rewriting of every ground constraint
$\leftarrow B(X)$ in $P$ with a rule $p(X) \leftarrow B(X), \neg
p(X)$, there is no recursive atom $A$ in $ground(P)$ depending on
itself through an odd number of negations. The formal proof of
this is outside the scope of this paper, and it could be
investigated in some future work. Another interesting problem to
be investigated in the future could be the syntactic
characterization of programs for which stable and strongly founded
models coincide. Clearly, this class contains positive and normal
programs and programs where head disjunctions are forced to be
exclusive by constraints.

%====================================================================0
%====================================================================0
%====================================================================0

\section{Conclusion}\label{conclusion}

The semantics proposed in this paper is essentially a variant of
stable model semantics for normal programs. The aim of our
proposal is the solution of some drawbacks of disjunctive stable
model semantics which, in some cases, interpret inclusive
disjunction as exclusive disjunction.

As disjunction is not
interpreted as exclusive, the proposed semantics is not invariant
if  rules which are subsumed by other rules (under stable model semantics)
are removed from the program; for instance, the first rule in the program
of Example \ref{Stratified-Example} can be deleted under stable
model semantics as it is subsumed by the second rule, whereas under the minimal
founded model semantics it cannot be deleted.

Several questions which need further investigation have been left open.
For instance, further research could be devoted to i) the
identification of fragments of disjunctive datalog for which one
minimal founded model can be computed in polynomial time; ii) the
use of two different types of disjunctive rule (inclusive
disjunction and exclusive disjunction), iii) the investigation of
abstract properties for disjunctive datalog under minimal founded
semantics \cite{BraDix95}.

\section*{Acknowledgement}

The authors are grateful to the anonymous referees for their
useful comments and suggestions.

\bibliography{tplp02}

\label{lastpage}

\end{document}